\documentclass[10pt,conference]{IEEEtran}
\usepackage{graphicx}
\usepackage{listings}
\usepackage{url}
\usepackage{subfigure}
\usepackage{textcomp}
\usepackage{amsmath}
\usepackage{array}
\usepackage{booktabs}
\newcolumntype{L}[1]{>{\raggedright\let\newline\\\arraybackslash\hspace{0pt}}m{#1}}
\newcolumntype{C}[1]{>{\centering\let\newline\\\arraybackslash\hspace{0pt}}m{#1}}
\newcolumntype{R}[1]{>{\raggedleft\let\newline\\\arraybackslash\hspace{0pt}}m{#1}}

\ifCLASSINFOpdf
 
\else
  
\fi

\usepackage{url}
\usepackage[ruled, linesnumbered]{algorithm2e}
\usepackage{color}

\begin{document}

\title{Towards Smart e-Infrastructures, A Community Driven Approach Based on Real Datasets}

\author{\IEEEauthorblockN{Prashant Singh \IEEEauthorrefmark{1}, Mona Mohamed Elamin \IEEEauthorrefmark{1}  and Salman Toor\IEEEauthorrefmark{1}}\\
\IEEEauthorblockA{\IEEEauthorrefmark{1} Department of Information Technology, Scientific Computing Division, Uppsala University, Sweden \\
Email: Mona.Mohamedelamin.6258@student.uu.se, \{Prashant.Singh, Salman.Toor\}@it.uu.se}
}



\maketitle

\begin{abstract}
e-Infrastructures have powered the successful penetration of e-services across domains, and form the backbone of the modern computing landscape. e-Infrastructure is a broad term used for large, medium and small scale computing environments. The increasing sophistication and complexity of applications have led to even small-scale data centers consisting of thousands of interconnects. However, efficient utilization of resources in data centers remains a challenging task, mainly due to the  complexity of managing physical nodes, network equipment, cooling systems, electricity, etc. This results in a very strong carbon footprint of this industry. In recent years, efforts based on machine learning approaches have shown promising results towards reducing energy consumption of data centers. Yet, practical solutions that can help data center operators in offering energy efficient services are lacking. This problem is more visible in the context of medium and small scale data center operators (the long tail of e-infrastructure providers). Additionally, a disconnect between solution providers (machine learning experts) and  data center operators has been observed. 
This article presents a community-driven open source software framework that allows community members to develop better understanding of various aspects of resource utilization. The framework leverages machine learning models for forecasting and optimizing various parameters of data center operations, enabling improved efficiency, quality of service and lower energy consumption. Also, the proposed framework does not require datasets to be shared, which alleviates the extra effort of organizing, describing and anonymizing data in an appropriate format.

\end{abstract}

\section{INTRODUCTION}
\label{INTRODUCTION}


e-Infrastructures have fueled substantial growth in the field of information technology, and enabled innovation of novel applications in the form of mobile apps and web-based services. At the heart of these cloud-based services lies efficient operation of massive data centers. Organizations including Google, Facebook, Amazon, Microsoft, as well as small and medium enterprises (SMEs), have built large-scale data centers in various parts of the world in order to offer efficient, robust, and affordable platforms and services based on reliable interconnects. 

In 2016, the size of this industry was approximately US \$219 billion, and by 2020 it is expected to grow to US \$411 billion \cite{forbes_forecast}. These are exciting times for the information technology industry, and market trends show that the growth of distributed computing infrastructures will continue. However, data centers must operate in energy and cost-efficient manner while maintaining hardware health and the computation performance of clusters. Balancing these conflicting requirements is challenging but critical for building and maintaining the next generation of smart e-infrastructures. A typical data center requires $538–-2,153$ $W/m^2$ on average and up to $10$ $kW/m^2$ in special cases \cite{avgerinou2017trends}. Cooling mechanisms alone contribute on average $40\%$ of the total power budget of a data center \cite{ni2017review}. In order to support business growth while maintaining the low energy consumption, a data center must exploit all available avenues, ranging from alternative energy sources to mechanisms for optimizing resource utilization.

Machine learning (ML) models have shown promising results in the area of data center research. Models for anomaly detection, resource allocation, and cyber-attacks have proved to be effective and are currently deployed in production environments. In recent years, a number of articles have been published in the area of modeling energy estimates, including highly accurate models that can predict energy consumption with an error of only $0.4\%$ \cite{42542}. The growing literature and encouraging results reflect the popularity and strength of the available solutions. Established organizations like Google are already using ML models for optimizing resource usage in data centers. The Deep Brain project reported $40\%$ better efficiency using ML models within their planning and configuration framework.

However, these remarkable results emanate from resourceful large-scale organizations who can invest in and build teams of experts. On the other hand, an overwhelming majority of institutions and research center have computing resources or own a managed facility (externally provided computing facilities). We refer to them as the ``Long tail of e-infrastructure providers" group and cumulatively they are significant in terms of resource utilization and energy consumption. The challenge for them remains the same, namely how to operate the infrastructure in an energy efficient manner while reducing the overall operating cost. This article presents our ongoing efforts to create a community-driven platform where infrastructure providers get the opportunity to use state-of-the-art ML models while maintaining a very low barrier to entry in terms of prerequisite ML knowledge.

The framework leverages the insight provided by machine learning models to answer questions considered to be the common denominator across data centers. This provides community members with a common denominator to better understand the results and learn by sharing knowledge and expertise. The article presents the following key contributions of this project: 

\begin{itemize}

\item \textbf{Open source framework}: a distributed framework that empowers operators to better understand infrastructure parameters and performance using ML models.


\item \textbf{Privacy preserved settings}: a framework that offers privacy-preservation while working within a community environment.

\item \textbf{ML solutions for practitioners}: a containerized approach that does not require infrastructure providers to have machine learning expertise. 

\end{itemize} 


The article is structured as follows. Section \ref{Machine learning for Computing Infrastructures} discusses machine learning efforts in literature towards aiding infrastructure operations. Section \ref{Challenges for Infrastructure Provides} introduces the important recurrent challenges faced by infrastructure providers. The framework architecture, including details on the mechanisms enabling answering high-level questions using machine learning models is explained in Section \ref{Framework Architecture}. 
Section \ref{Community Members and datasets} presents community-wide ongoing efforts. Section \ref{Results and Discussion} highlights our approach and the findings based on real datasets. Finally, Section \ref{CONCLUSION} concludes the paper.

\section{Machine learning for Computing Infrastructures}
\label{Machine learning for Computing Infrastructures}


Machine learning has been applied towards improving various aspects of data center operation. Frameworks have been proposed in the past that involve turning on or off certain machines and use power-aware consolidation algorithms to achieve green(er) computing \cite{berral2010towards}. Google have leveraged vast amounts of historical data usage logs from various data centers to train a highly accurate neural network model that can predict energy consumption based on ongoing usage patterns \cite{42542}. Indeed, the use of machine learning is very popular for tasks such as energy consumption modeling \cite{dayarathna2016data}, resource provisioning \cite{zhang2016resource} and scheduling \cite{chen2015towards}, optimizing cooling systems \cite{zhang2016towards}, monitoring and preventing service level agreement (SLA) violations \cite{leitner2010monitoring}, etc.



Although there exists rich literature concerning the use of machine learning approaches towards improving data center performance and operation, there is a disconnect between academic research and industry adoption. Even though large organizations like Google have investigated and incorporated machine learning approaches in various aspect of data center operation \cite{42542}, the scope of such efforts in practicality has been limited within organizations. This is partly due to an entry barrier presented by the lack of ML expertise in the mass market. This article aims at addressing this issue by proposing a Docker-based easy to deploy framework, as outlined below.


\begin{figure*}[t]
  \caption{A distributed framework architecture based on three important pillars Publisher, Subscriber and community's repository.}
  \centering
    \includegraphics[width=18cm,height=6cm]{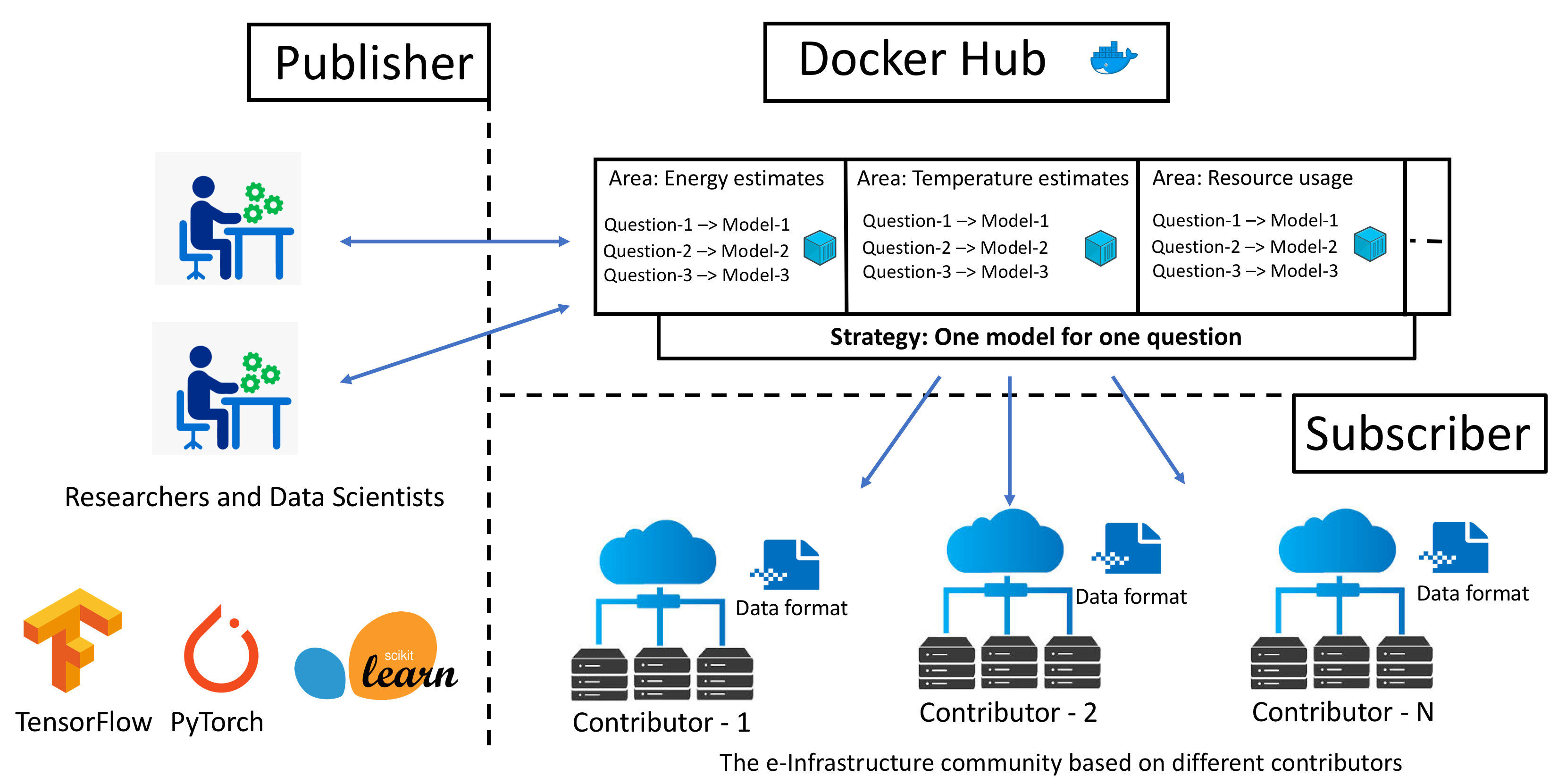}
     \label{fig:ei-architecture}
\end{figure*}

\section{Challenges for Infrastructure Providers}
\label{Challenges for Infrastructure Provides}

Currently, most infrastructure providers still use monitoring based alarms to offer $24 / 7$ services. We have found the following major reasons for the gap between the solutions designed by data scientists and the operational approaches to run infrastructure services:

\begin{itemize}

\item \textbf{Lack of expertise}: for small and medium scale organizations, it is often not feasible to dedicate full-time personnel to develop complex machine learning models. It is a formidable initial barrier to overcome as it is often not an area of expertise of e-infrastructure providers. 

\item \textbf{Studies are based on synthetic datasets}: often the presented results are based on simulated or composed datasets. Sometimes datasets are either too old or specialized which makes it difficult to relate with local environment settings. 

\item \textbf{Missing common denominator}: developing a basic understanding with publicly available datasets or datasets from other organizations is time-consuming and it is often not clear how to relate results with local environment settings. On the other hand, a number of different machine learning models are available and it is not easy to discuss results without having a firm understanding of those models. 

\item \textbf{The studies are ML focused}: most of the available literature highlights the challenges solved within the machine learning domain by presenting field specific datasets as use-cases. Performance, efficiency and accuracy of the ML models constitute the main focus of such studies. This often leads to unanswered actual domain specific questions while focusing too intensely on approach-specific details.

\item \textbf{Privacy and security}: in some cases sharing data is not an option as it can raise privacy concerns and can potentially make infrastructures vulnerable to security breaches. 

\item \textbf{Procurement of new hardware}: another recurring challenge for infrastructure providers is the purchase of new hardware. The hardware purchase is a critical process and it requires firm understanding of the hardware and the application workloads. A community based approach where contributors share their experiences based on common denominators can significantly help in this process.        

\end{itemize} 

Based on the aforementioned challenges, we have initiated a community-driven approach enabling answers to practical questions consisting of on-premises familiar settings, alleviating the need for data sharing. The proposed open framework can also be considered as a source of information based on facts, experiences and shared knowledge that helps operators in selecting the best possible equipment within the available budget. 

At the heart of this platform lie powerful machine learning models that can provide comprehensive insights about the infrastructure under study. This can only be possible if the same set of ML models are accessible to every contributor, with a set of consistent usage guidelines and format. Here it is important to note that with the proposed approach, operators do not require to develop an understanding of the complex modeling details. The machine learning aspect is a black-box which accepts input data and generates the output explaining the underlying behaviors based on different parameters.

\section{Framework Architecture}
\label{Framework Architecture}

The proposed architecture is based on a loosely coupled model where contributors have complete autonomy over their respective infrastructure datasets. The architecture is based on following three parts:

\begin{itemize}
    \item \textbf{Publisher}: The data science community that develops various models and wants to test the strength of their proposed solutions. They typically use standard libraries such as TensorFlow \cite{tensorflow}, PyTorch \cite{pytorch} and scikit-learn \cite{scikit}. However, the proposed architecture requires that models be packaged as software containers and community members need not concern themselves with the complexities lying within the containers. 
    \item \textbf{Subscriber}: The e-infrastructure providers with the goal of optimally operating their resources. They pose questions and \textit{Publishers} build models to answer those questions and package them as software containers. The \textit{Subscribers} need to download the container and run it with the locally available datasets.   
    \item \textbf{Open-source platform for model sharing}: In order to make containerized models available to the community members, we use the Docker container technology and Docker-Hub for container availability. This choice allows us to use public, private and locally hosted container repositories.
\end{itemize}

The framework does not require community members to pool data in a shared location. Instead of data from contributors moving to a central location for training, the machine learning models as docker containers travel towards datasets. Another advantage of proposed architecture is that the \textit{Publishers} can access raw containers and train the models based on the local datasets. However, based on mutual trust, community members can also jointly train models and capture the behaviors that currently lie outside to their infrastructures, but could be of interest. Here it is important to note that in both cases the datasets are still not required to be shared even between trusted community members. 

\begin{table*}[h]
\scriptsize
\centering
\caption{Common denominator based on one-model one-question approach. Time Resolution: Daily/Weekly/Monthly}
\label{table:framework}
\begin{tabular}{C{6.5cm}|C{3cm}|C{6.5cm}}
\toprule
Question & Model & Description \\
\midrule
\multicolumn{3}{c}{Node Level}\\
What is the minimum, average and maximum energy consumption of a particular node? (E.g., in case of mission critical specialized equipment.) & $\hat{y}_p^i = m_p^i({\bf x})$ & A ML model capturing the relationship between parameters of node $i$ ${\bf x}$ and predicted power consumption ${\hat{y}_p}.$ \\
\midrule
\multicolumn{3}{c}{Cluster Level}\\
What is the minimum, average and maximum energy consumption of a cluster? & $\hat{y_c}_p = \sum_{i=1}^K \hat{y}_p^i$ & The predictions from $K$ node-level ML models can be aggregated to obtain cluster-level predictions for a $K$-node cluster. At a more advanced level, node-level ML models can be combined with cluster level parameters ${\bf x}_c$ to train more intricate, detailed models. \\
\midrule
\multicolumn{3}{c}{Resource Usage}\\
What is the relationship between energy consumption, and CPU, memory, disk and network utilization? & $m_p^i({\bf x})$, $m_c^i({\bf x}_1)$, $m_m^i({{\bf x}_2})$, $m_d^i({\bf x}_3)$, $m_p^i({\bf x}_4)$ & Let ${\bf x}$ be the parameters affecting node-level energy consumption and $\{{\bf x}_j\}_{j=1}^4$ be independent variables (parameters) affecting node-level models for CPU $m_c$, memory $m_m$, disk $m_d$ and network $m_n$ usage respectively. Time series regression models with exogenous inputs (${\bf x}$'s) corresponding to each quantity of interest can be trained for forecasting and planning ahead of time. \\
\midrule
\multicolumn{3}{c}{Temperature Estimates}\\
What are the average and daily temperature estimates? & $\hat{y}_t^i = m_t^i({\bf x})$ & A ML model capturing the relationship between parameters of node $i$ ${\bf x}$ and predicted node temperature ${\hat{y}_t}.$ \\
\midrule
\multicolumn{3}{c}{Network Level}\\
What is the minimum, average and maximum energy consumption of network devices? & $\hat{y}_n^i = m_n^i({\bf x})$ & A ML model capturing the relationship between parameters of node $i$ ${\bf x}$ and predicted network energy consumption ${\hat{y}_n}.$ \\
\bottomrule
\end{tabular}
\end{table*}

Table \ref{table:framework} presents examples of specific high level questions, formulation and description of corresponding models, and the desired time resolution within the context of the question. As an example, model $m_p^i$ is a machine learning model trained to learn the variation in energy consumption patterns for a given node $i$, based on specific parameters or variables ${\bf x}$. The variables ${\bf x}$ may include quantities such as CPU, memory and disk utilization. 

Similarly, other models corresponding to specific distinct questions can be trained at either the node-level or even cluster-level. The node-level models can be leveraged and aggregated to obtain insights on cluster-level behaviors. As as example, Table \ref{table:framework} lists a question concerning cluster-level relationship between energy consumption. The model approximation $\hat{y_c}_p = \sum_{i=1}^K \hat{y}_p^i$ aggregates node-level approximations to obtain cluster-level insight.

Table \ref{table:framework} also lists a question concerning resource usage. Namely, the relationship between energy consumption of a node, and quantities CPU, memory, disk and network usage. Each of these quantities may depend on distinct independent variables, namely ${\bf x}$ and $\{{\bf x}_j\}_{j=1}^4$. The corresponding trained models $m_p^i({\bf x})$, $m_c^i({\bf x}_1)$, $m_m^i({{\bf x}_2})$, $m_d^i({\bf x}_3)$, $m_p^i({\bf x}_4)$ can be used for obtaining highly detailed forecasting of each quantity, as well as for studying the fine effects each quantity has on node-level energy consumption. This and models predicting temperature estimates can be a powerful tool for planning and forecasting, including for tasks such as hardware upgrade windows, cooling system maintenance, financial forecasting and planning, etc.


The success of this framework is based on packaging of models in a way that insulates contributors from complex internal details. It is important to consider a container as a black-box that requires path to a directory containing input files based on the defined format and returns another path of the directory containing the results.

Based on the proposed methodology, we have built the framework using open sources tools. The containers based on the questions are available via the DockerHub public repository [https://github.com/sztoor/smart-infrastructures].

\section{Community Members and datasets}
\label{Community Members and datasets}
There has been encouraging interest from various organizations towards early adoption of the proposed platform. The organizations include CSC - IT center for science Finland \cite{csc}, UPPMAX - Uppsala Multidisciplinary Center for Advanced Computational Science Sweden \cite{uppmax} and C3SE - Chalmers Center for Computational Science and Engineering Sweden \cite{c3se}. We are also in touch with other infrastructure providers and hope to grow the community base in the near future. The presented results are based on datasets generated by the CSC and UPPMAX resources. We are also very grateful to our community members for their support to formulate pertinent questions mentioned in the Table \ref{table:framework}. The presented results are based on the questions related to energy consumption, machine level temperature readings and resource (CPU, memory, network etc) utilization.  

The collection and organization of local datasets requires information about acceptable data formats. The list of supported data formats will be drawn up in consultation with the community members such that all variables affecting important questions are captured in the format. The choice also depends on questions of interest, data size and collection mechanism. This work is ongoing and future publications will shed more light on data formats of interest. Currently, a standard CSV (comma-separated-value) format is followed with predefined parameters. An example is presented in the following section.

\section{Results and Discussion}
\label{Results and Discussion}

The results presented in this section are organized with respect to the domain specific questions. This approach highlights our commitment with strong focus towards e-infrastructure related questions rather then the ML modeling methods. 
The presented results show reliable predictions based on available information which leads to energy efficient and cost effective solutions. Here it is important to note that the presented results in this section are based on real datasets collected at CSC and UPPMAX data centers.

For the purpose of experiments, the machine learning models used include neural network regression models \cite{haykin2009neural} and long short-term memory (LSTM) time series models. 
LSTM models constitute a variant of recurrent neural networks (RNNs) and have proven to be effective towards capturing long term temporal variations and dependencies across diverse application domains \cite{greff2016lstm}. LSTMs have the ability to learn longer sequences of patterns or states by virtue of memory blocks. 
A thorough discussion of LSTMs is out of scope of this work, and the reader is referred to \cite{greff2016lstm} for a deeper treatment.

\textit{Q-1: What is the predictive CPU utilization of mission critical machines? }

\begin{figure}[h]
\centering
\subfigure[$CSC-machine1$.]{
\label{fig:m1_lstm}
\includegraphics[width=0.35\textwidth]{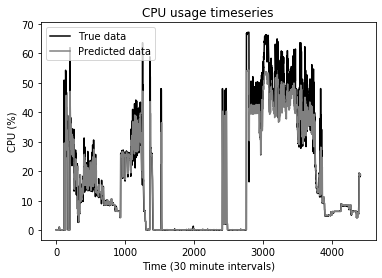}}
\subfigure[$CSC-machine2$.]{
\label{fig:m2_lstm}
\includegraphics[width=0.35\textwidth]{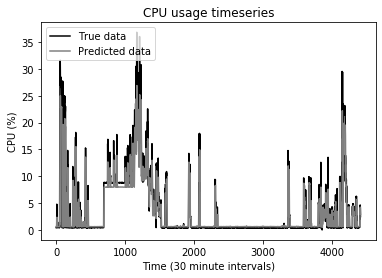}}
\caption{Actual and predicted CPU series}
\end{figure}

Figure \ref{fig:m1_lstm} and \ref{fig:m2_lstm} show the CPU time-series prediction for two different machines with LSTM recurrent neural networks. For the experiments in this work, initial $70\%$ of the data is used for training and validation, while the rest is used for testing. Therefore, the final $30\%$ of the time series in the figures is predicted using the models.
The prediction mean absolute error (MAE) on the test set for $CSC-machine1$ is $2.96$ and mean squared error (MSE) is $30.42$. This indicates a reasonably accurate model that is only about $3\%$ off true values on average. The error is on average very low throughout, except for in between intervals $3000$ and $4000$ where highly varied non-linear responses are present.
On the other hand, the model for $CSC-machine2$ has a MAE of $1.47$ and MSE of $2.96$ on the test set.  In Fig. \ref{fig:m2_lstm} it can be seen that the model is very accurate over majority of the time intervals. The error is predominantly in the two peaks corresponding to high CPU usage. This model is useful for CPU forecasting and to better understand the CPU utilization over time. The provided information helps to design effective and reliable services consolidation\footnote{Services consolidation is a well-know approach for efficient resource utilization.} scheme in data centers.  

\textit{Q-2: What is the predictive power consumption based on CPU and Network usage?}

\begin{figure}[h]
  \centering
  \includegraphics[width=0.4\textwidth]{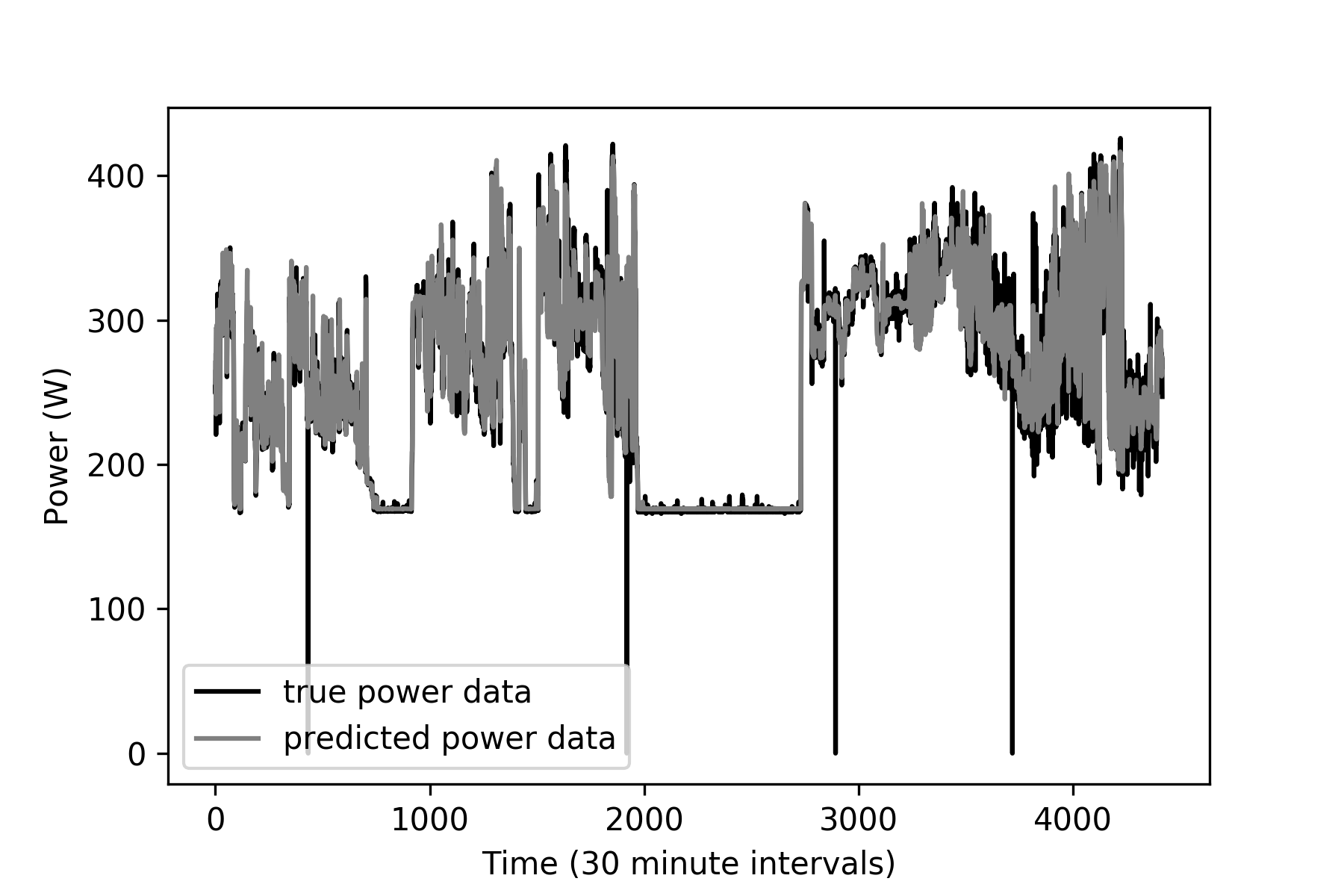}
  \caption{Power prediction using CPU and Network.}
\label{fig:powerCpuNet}
\end{figure}


Figure \ref{fig:powerCpuNet} shows the power consumption prediction of a machine with power values ranges from $0$ to $348$ Watts. CPU usage and network usage are the input features that the model depends on to predict the power consumption. MSE on the test set is approximately $105$, which appears to be high but is amplified by the presence of four outliers (machine being rebooted or monitoring system failed to capture the readings). It can be observed from Fig. \ref{fig:powerCpuNet} that the model is accurate throughout the time intervals barring the outliers. This model is useful to estimate the power consumption, and can be used while assigning tasks to machines based on expected CPU demands. Additionally, it can be used to limit the power values by limiting the corresponding values of CPU and network.

\textit{Q-3: What is the relationship between power consumption and CPU and Network usage?}

\begin{figure}[h]
  \centering
  \includegraphics[width=0.4\textwidth]{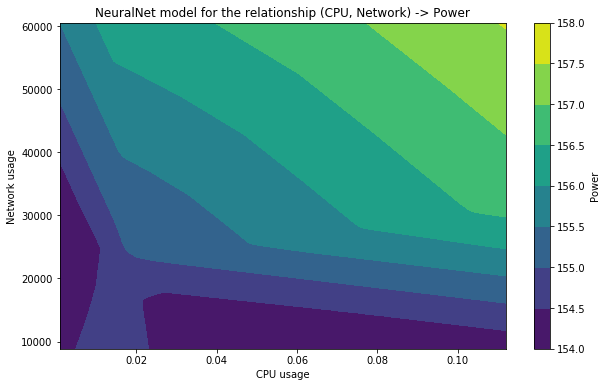}
  \caption{Power prediction using CPU and Network.}
\label{fig:e101}
\end{figure}

Figure \ref{fig:e101} depicts a model highlighting the relationship of power consumption with respect to varying CPU and network usage, irrespective of time. The contour plot consists of CPU usage represented along the x-axis, and network usage represented along the y-axis. The varying colours correspond to varying levels of power consumption listed in the color-map corresponding to the figure. 


 For low values of CPU and network usage, the power consumption is low. The power consumption grows as network and CPU usage increase. The figure and in turn the corresponding model, although simple and intuitive, can be a very powerful tool for inferring limits of CPU and network usage according to a known or desired power consumption level. As an example, the CPUs can be throttled back to a specific value in order to meet monthly power consumption budgets.


\textit{Q-4: What is the predictive temperature on different machines based on resource (CPU, memory, Network, Disk IOs) utilization?}


\begin{figure}[h]
\centering
\subfigure[$SSC-machine1$.]{
\label{fig:uppM1Result}
\includegraphics[width=0.4\textwidth]{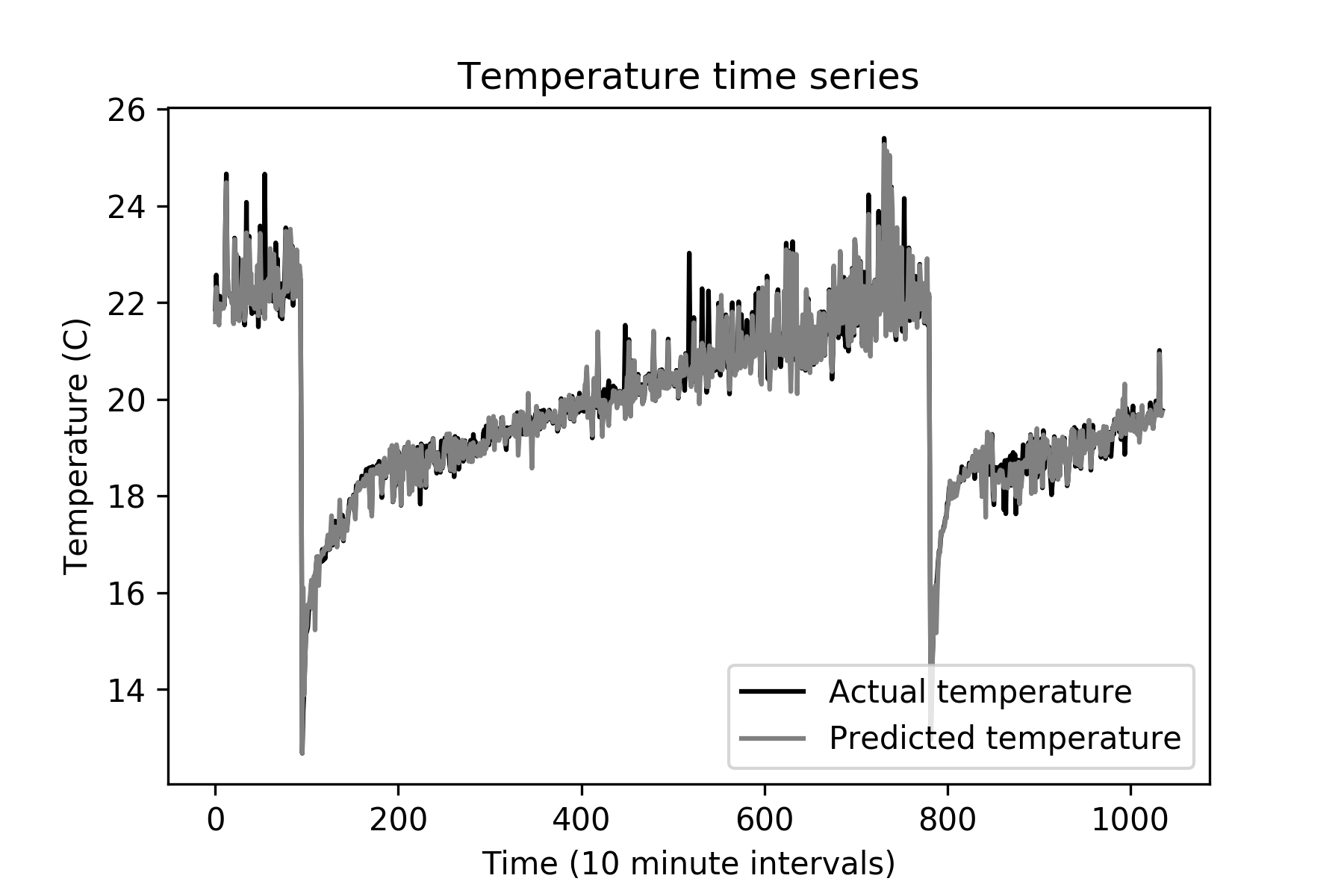}}
\subfigure[$SSC-machine2$.]{
\label{fig:uppM2Result}
\includegraphics[width=0.4\textwidth]{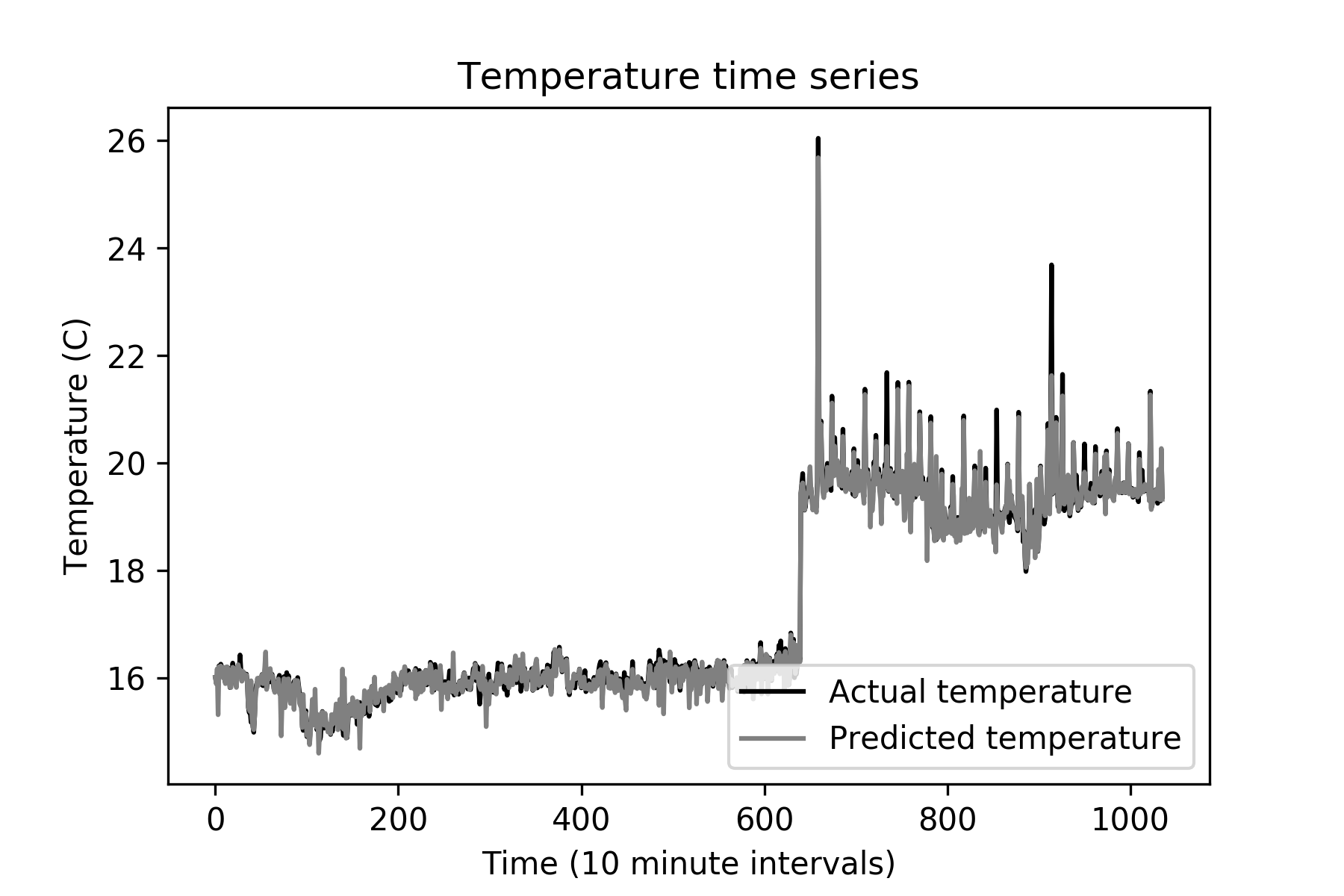}}
\caption{Temperature predictions using multiple features (CPU, memory, disk I/O, and Network) }
\end{figure}

Figure \ref{fig:uppM1Result} and \ref{fig:uppM2Result} show the temperature predictions for two different machines. 
Input features used to predict temperature are CPU, memory, disk input/output, and disk space usage. MSE error on a test set for $SSC-machine1$ is $0.014$ and $0.005$ for $SSC-machine2$. Both values indicate highly accurate models. 

As mentioned in Section \ref{Challenges for Infrastructure Provides}, procurement and placement of new hardware are recurring challenges for the data center operators. The provided temperature estimates can be very useful both for the placement and procurement of new hardware in a data center.



\section{CONCLUSIONS}
\label{CONCLUSION}
This article presented a community-driven framework that is aimed at enabling insights into various aspects of e-infrastructure operations, helping operators in offering cost effective services while minimizing energy consumption. The article aims to bridge the gap between data scientists developing state-of-the-art machine learning solutions, and small and medium sized infrastructure operators who would benefit greatly from such solutions, but lack machine learning expertise.
The framework's architecture inherently provides privacy-preserving settings that does not require data sharing between different stake holders. It allows small and medium size infrastructure providers to pool together local knowledge towards realizing global patterns without sharing their data. The global knowledge is encapsulated within machine learning models that are delivered via Docker containers and do not require machine learning expertise to set up. The proposed framework is thus designed to make machine learning based solutions accessible and easy to setup and use within the context of smart e-infrastructures.

\section*{ACKNOWLEDGMENTS}
This work has been supported by the  Swedish strategic initiative for eScience, eSSENCE \cite{essence}. The computations were performed on resources provided by the SNIC Science Cloud (SSC)\cite{ssc}. We would also like to thanks UPPMAX \cite{uppmax} and CSC \cite{csc} for providing us the valuable datasets. 

\bibliographystyle{IEEEtran}
\bibliography{references} 

\end{document}